\documentclass[twocolumn,superscriptaddress,notitlepage]{revtex4-2}
\usepackage[utf8]{inputenc}
\usepackage[T1]{fontenc}
\usepackage[english]{babel}
\usepackage[dvipsnames]{xcolor}
\usepackage{lmodern,fancyhdr}
\usepackage{graphicx,float}
\usepackage{enumitem,amsmath,amssymb,amsthm,bm,array,mathdots,mathrsfs,dsfont,bbold}
\usepackage[colorlinks]{hyperref}
\hypersetup{colorlinks,linkcolor=red,citecolor=blue,urlcolor=blue,final}

\newcommand{\im}{\text{i}}
\newcommand{\hc}{{\rm H.c.}}
\newcommand{\pare}[1]{\left( {#1} \right)}
\newcommand{\spare}[1]{\left[ {#1} \right]}
\newcommand{\cpare}[1]{\left\{ {#1} \right\}}
\newcommand{\pares}[1]{( {#1} )}
\newcommand{\spares}[1]{[ {#1} ]}
\newcommand{\cpares}[1]{\{ {#1} \}}
\newcommand{\be}{\begin{equation}}
\newcommand{\ee}{\end{equation}}
\newcommand{\eqnref}[1]{Eq.~\eqref{#1}}
\newcommand{\figref}[1]{Fig.~\ref{#1}}
\newcommand{\secref}[1]{Sec.~\ref{#1}}

\newcommand{\fd}[2]{\frac{\text{d} {#1}}{\text{d} {#2}}}
\newcommand{\Hop}{\hat{H}}
\newcommand{\aop}{\hat{a}}
\newcommand{\aopd}{\hat{a}^\dagger}
\newcommand{\bop}{\hat{b}}
\newcommand{\bopd}{\hat{b}^\dagger}
\newcommand{\cop}{\hat{c}}
\newcommand{\copd}{\hat{c}^\dagger}
\newcommand{\Xop}{\hat{X}}
\newcommand{\mv}[1]{ \langle #1 \rangle }
\newcommand{\anglesq}{\theta_\text{sq} }
\newcommand{\varsq}{\Delta^2 X_\text{sq} }
\newcommand{\rhoop}{\hat{\rho}}

\newcommand{\Dopt}{\Delta_\text{opt}}
\newcommand{\tstar}{t^\star}
\newcommand{\Psopvec}{\hat{\bm{\Psi}}}
\newcommand{\Phopvec}{\hat{\bm{\Phi}}}
\newcommand{\pseudo}{\mathscr{I}}
\newcommand{\psop}{\hat{\psi}}

\newcommand{\phop}{\hat{\phi}}

\newcommand{\iop}{\hat{o}}
\newcommand{\iopd}{\hat{o}^\dagger}
\newcommand{\Sopvec}{\hat{\bm{S}}}

\newcommand{\Sop}{\hat{S}}
\newcommand{\afac}[1]{ a_{\scriptscriptstyle #1} }
\newcommand{\bfac}[1]{ b_{\scriptscriptstyle #1} }
\newcommand{\integral}[3]{ \int_{#1}^{#2} \text{d} {#3} \hspace{1mm} }

\begin{document}

\title{Mechanical squeezing via unstable dynamics in a  microcavity}
\author{Katja~Kustura}
\affiliation{Institute for Quantum Optics and Quantum Information of the Austrian Academy of Sciences, A-6020 Innsbruck, Austria.}
\affiliation{Institute for Theoretical Physics, University of Innsbruck, A-6020 Innsbruck, Austria.}

\author{Carlos~Gonzalez-Ballestero}
\affiliation{Institute for Quantum Optics and Quantum Information of the Austrian Academy of Sciences, A-6020 Innsbruck, Austria.}
\affiliation{Institute for Theoretical Physics, University of Innsbruck, A-6020 Innsbruck, Austria.}

\author{Andr\'es~de~los~R\'ios~Sommer}
\affiliation{Nanophotonic Systems Laboratory, Department of Mechanical and Process Engineering, ETH Zurich, 8092 Zurich, Switzerland.}
\affiliation{Quantum Center, ETH Zurich, 8083 Zurich, Switzerland.}

\author{Nadine~Meyer}
\affiliation{Nanophotonic Systems Laboratory, Department of Mechanical and Process Engineering, ETH Zurich, 8092 Zurich, Switzerland.}
\affiliation{Quantum Center, ETH Zurich, 8083 Zurich, Switzerland.}

\author{Romain~Quidant}
\affiliation{Nanophotonic Systems Laboratory, Department of Mechanical and Process Engineering, ETH Zurich, 8092 Zurich, Switzerland.}
\affiliation{Quantum Center, ETH Zurich, 8083 Zurich, Switzerland.}

\author{Oriol~Romero-Isart}
\affiliation{Institute for Quantum Optics and Quantum Information of the Austrian Academy of Sciences, A-6020 Innsbruck, Austria.}
\affiliation{Institute for Theoretical Physics, University of Innsbruck, A-6020 Innsbruck, Austria.}

\begin{abstract}
	We theoretically show that strong mechanical quantum squeezing in a linear optomechanical system can be rapidly generated through the dynamical instability reached in the far red-detuned and ultrastrong coupling regime. We show that this mechanism, which harnesses unstable multimode quantum dynamics, is particularly suited to levitated optomechanics, and we argue for its feasibility for the case of a levitated nanoparticle coupled to a microcavity via coherent scattering. We predict that for sub-millimeter-sized cavities the particle motion, initially thermal and well above its ground state, becomes mechanically squeezed  by tens of decibels on a microsecond timescale. Our results bring forth optical microcavities in the unresolved sideband regime as powerful mechanical squeezers for levitated nanoparticles, and hence as key tools for quantum-enhanced inertial and force  sensing.
\end{abstract}

\maketitle

A system of linearly coupled quantum harmonic oscillators can be dynamically unstable, even in the absence of dissipation (see~\cite{Kustura2019} and references therein). Although dynamically unstable regimes are often considered undesirable~\cite{Han2019}, they can also be a resource. In this article we show how to engineer and harness unstable multimode dynamics in a linear optomechanical system to induce strong mechanical squeezing. Preparing squeezed states of a mechanical oscillator -- quantum states with position uncertainty smaller than its zero-point motion -- is a key aim of optomechanics~\cite{Aspelmeyer2014}, as these states lie at the heart of many quantum-enhanced force and inertial sensing schemes~\cite{Caves1980}. This is evidenced by the many theoretical and experimental efforts aimed at developing strategies to generate mechanical squeezing, e.g. reservoir engineering based on two-tone driving~\cite{Kronwald2013,Tan2013,Kienzler2015,Wollman2015,Pirkkalainen2015,Lecocq2015,Lei2016}, parametric squeezing~\cite{Milburn1981,Rugar1991,Blencowe2000,Mari2009,Nunnenkamp2010,Liao2011,Vinante2013,Pontin2014,Pontin2016,Chowdhury2020,Vezio2020}, rapid frequency shifts~\cite{Janszky1986,Lo1990,Janszky1992,Asjad2014,Alonso2016,Rashid2016}, quantum measurement~\cite{Clerk2008, Szorkovszky2011,Meng2020,Filip2005,Vanner2011,Szorkovszky2013,Vanner2013,Vasilakis2015}, mechanical nonlinearities~\cite{Nunnenkamp2010,Lu2015,Xiao2017,Asjad2014}, or quantum transfer of a squeezed state from a cavity mode to the mechanical oscillator~\cite{Jahne2009,Agarwal2016}.

In this article, we propose a novel approach based on the fast unstable quantum dynamics of a linear optomechanical system. Our protocol requires to operate in the ultrastrong coupling and far red-detuned regime, i.e. $g > \Omega/2$ and $\Delta \gg \Omega$ with $\Omega$ the mechanical frequency, $\Delta$ the laser detuning from cavity resonance, and $g$ their optomechanical coupling rate [see \eqnref{Hamiltonian}]. This regime is within reach for levitated optomechanics~\cite{Millen2020,GonzalezBallestero2021}, specifically for an optically levitated nanoparticle coupled via coherent scattering~\cite{Windey2019,Delic2019,Delic2020,Sommer2021,Ranfagni2021} to a microcavity~\cite{Hunger2010,Magrini2018,Wachter2019,Fait2021} in the unresolved sideband regime. Our results are particularly timely due to recent experiments demonstrating ground-state cooling and quantum control of optically levitated nanoparticles in free space~\cite{Magrini2021,Tebbenjohanns2021}. As opposed to the first ground-state cooling experiment~\cite{Delic2019}, which required a resolved-sideband cavity —  and thus squeezing protocols designed for such regime~\cite{Cernotik2020} —, these recent free-space ground-state cooling experiments do not need an optical cavity. Our approach thus allows us to incorporate an independent and passive {\em mechanical squeezer} to these state-of-the-art experiments in the form of a properly optimized microcavity. The sole purpose of such a cavity is the generation of strong squeezing in the motion of the cooled levitated nanoparticle. This opens the door toward achieving quantum-enhanced sensing with levitated microobjects~\cite{Millen2020,Moore2021,GonzalezBallestero2021}.

We consider a mechanical oscillator of mass $m$ and frequency $\Omega$ coupled to an optical cavity mode of frequency $\omega_c$ that is being driven at frequency $\omega_t$ in the red-detuned regime $\Delta \equiv \omega_c - \omega_t >0$. The linearized Hamiltonian of the system in a frame rotating at the frequency of the driving is given by
\be \label{Hamiltonian}
\begin{split}
	\Hop =  \hbar\Delta\aopd \aop + \hbar\Omega \bopd\bop + \hbar g \pares{\aopd+\aop}\pares{\bopd+\bop}, 
\end{split}
\ee 
where $\aop$ and $\bop$ are bosonic annihilation operators of the cavity mode and the mechanical mode, respectively. In this article we focus solely on the regime~\cite{Genes2008}
\be \label{unstable}
\begin{split}
	\frac{4 g^2}{\Delta \Omega} > 1,
\end{split}
\ee 
which makes the system described by~\eqnref{Hamiltonian} dynamically unstable~\cite{Peterson2019}. In the unstable  regime defined by \eqnref{unstable}, the Hamiltonian in \eqnref{Hamiltonian} cannot be diagonalized in terms of bosonic modes, but it can be expressed in normal form~\cite{Kustura2019} as
\be \label{Hamiltonian-nf}
\begin{split}
	\Hop = \hbar\omega_1 \hat c^\dagger_1 \hat c_1  + \frac{\im \hbar r}{2}\spare{(\hat c_2^\dagger)^2-\hat c_2^2},
\end{split}
\ee 
where $\hat c_1$ and $\hat c_2$ are bosonic annihilation operators of the normal modes, $\omega_1^2 \equiv ({\zeta^2}+\Delta^2+\Omega^2)/2$, $r^2 \equiv ({\zeta^2}-\Delta^2-\Omega^2)/2$, and $\zeta^4 \equiv (\Delta^2-\Omega^2)^2+ 16 \Delta \Omega g^2$. The canonical transformation between the physical modes~$\{\hat a$, $\hat b\}$ and the normal modes~$\{\hat c_1$, $\hat c_2\}$ is given in~\cite{supplementary}. The Hamiltonian in~\eqnref{Hamiltonian-nf} elucidates the dynamics in the unstable regime: the normal mode $\hat c_1$ is described by an uncoupled harmonic oscillator term, and the normal mode $\hat c_2$  by a pure squeezing term with a squeezing rate $r$ that accounts for the unstable dynamics of the system. A key observation  is that in the far-detuned regime $\Delta \gg \Omega$, the squeezed hybrid mode $\hat c_2$  is dominated by the contribution of the mechanical mode, namely
\be
\lim_{\Delta/ \Omega \gg 1}\hat c_2 = - \im \frac{g}{\sqrt{\Omega \Delta}} \spare{ \pares{\bopd+\bop} + \im \sqrt{\frac{\Omega }{\Delta}}\pares{\aopd-\aop}   }.
\ee 
This indicates that mechanical squeezing should be dynamically generated if, in addition to the instability condition defined by \eqnref{unstable}, the condition $\Delta \gg \Omega$ (far red detuning) is fulfilled. Note that both requirements can only be satisfied in the ultrastrong coupling regime $g \gg  \Omega/2$~\cite{Anappara2009,Bayer2017,FriskKockum2019,Fogliano2021}. One can show that in the far red-detuned regime the squeezing rate in \eqnref{Hamiltonian-nf} is given by $r \approx 2 g \sqrt{\Omega/\Delta}$. In the following we show how mechanical squeezing is generated. 

\begin{figure}[t]
	\centering
	\includegraphics[width=1\linewidth]{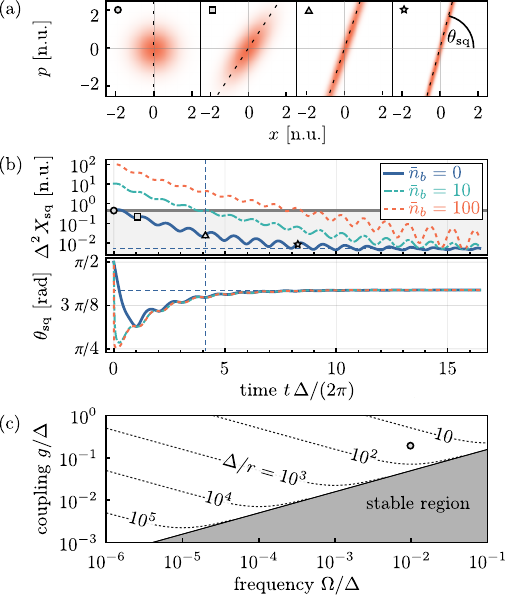}
	\caption{ Squeezing induced via dynamical instability in the absence of dissipation. 
		(a)~Wigner function of the mechanical state $W(x,p)$ at different times [denoted by points in panel (b)], for an initial state with mean phonon number $\bar n_b = 0$.  
		(b)~Minimal variance $\varsq$ and the angle of squeezing $\anglesq$ as a function of time, for $g/\Delta=0.2$, $\Omega/\Delta=0.01$, for various initial thermal states with mean phonon number $\bar n_b$, and for a cavity initially in the vacuum state. Dashed vertical and horizontal lines show the squeezing timescale $\Delta/r$ and the asymptotic value $\Omega/(2 \Delta)$, respectively. Gray shaded area indicates squeezing. 
		(c)~Squeezing timescale $\Delta/r$ as a function of the mechanical frequency $\Omega$ and the optomechanical coupling rate $g$. The point denotes the configuration analyzed in panels (a) and (b). 
	}
	\label{fig:unstable_dynamics}
\end{figure}

Let us define the generalized mechanical quadrature $\Xop(\theta)  \equiv \pares{ \bopd e^{\im \theta}   + \bop e^{-\im \theta}  }/\sqrt{2}$, with $\theta \in [0,2\pi)$. The minimal variance is obtained at a phase-space angle $2 \anglesq \equiv \arg{ \mv{\bop^2} }$ [see \figref{fig:unstable_dynamics}(a)] and is given by $\varsq \equiv  \mv{\Xop^2(\anglesq + \pi/2)} = 1/2 + \mv{\bopd\bop} - \vert\mv{\bop^2}\vert$. Squeezed states are defined by a variance $\varsq < 1/2$  and their squeezing is quantified in decibels (dB) by $S \equiv -10 \log_{10}\pares{ 2\varsq }$. We consider the coherent dynamics generated by \eqnref{Hamiltonian} with an initial state given by the cavity mode in vacuum (in the linearized regime) and the mechanical mode in a thermal state with mean phonon number $\bar n_b $~\footnote{We remark that the conservative dynamics generated by~\eqnref{Hamiltonian} and discussed in~\figref{fig:unstable_dynamics} can be understood in terms of the approximated exponentiation discussed in~\cite{Brandao2021}}. In \figref{fig:unstable_dynamics}(b) we show $\varsq$ and $\anglesq$ as a function of time, thereby demonstrating the generation of mechanical squeezing. One can show that the variance $\varsq$ reaches the asymptotic value 
\be \label{AsymptoticX}
\lim_{t r \gg 1} \varsq = \frac{1}{2} \frac{\Omega}{ \Delta} \ll 1.
\ee
Remarkably, the asymptotic value is independent of the mean phonon number $\bar n_b$. Squeezing is achieved at a phase-space angle given by  $  \lim_{t r \gg 1}  \exp[2 \im \anglesq] \approx  -1 +  \Delta \Omega/(2g^2) + \im \sqrt{\Delta \Omega}/g$. Figure \ref{fig:unstable_dynamics}(c) shows the squeezing timescale $\Delta/r$ throughout the stability diagram, considering the exact expression for $r$, which shows deviation from the approximated expression $r \approx 2 g \sqrt{\Omega/\Delta}$ close to the stability border. If the asymptotically squeezed state with the variance given in \eqnref{AsymptoticX} is rotated in phase space such that the position quadrature $\hat X (\theta=0)$ becomes maximally squeezed, the corresponding position fluctuations are given by $\sqrt{\hbar/(2 m \Delta)}$. These fluctuations are similar to the zero-point motion associated with a hypothetical harmonic trap with frequency $\Delta \gg \Omega$. Such rotation in phase space can be done via free evolution in the harmonic trap for a time $\Omega t = \anglesq +\pi/2$. 

\begin{figure}[t]
	\centering
	\includegraphics[width=1\linewidth]{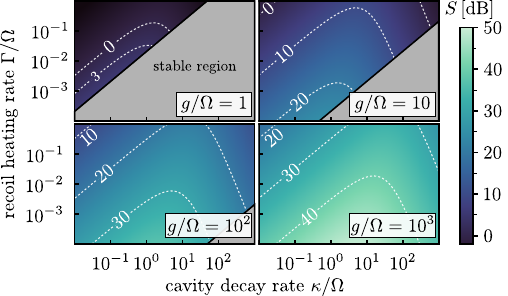}
	\caption{Asymptotic mechanical squeezing $S$ in the presence of dissipation, considering optimal detuning $\Delta = \Delta_\text{opt}$ (see main text), and an initial state with  mean phonon number $\bar n_b = 0$.  }
	\label{fig:dissipation}
\end{figure}

Let us now discuss how the presence of noise and decoherence affects these results. We  consider the dynamics described by the following master equation,
\begin{multline}  \label{master-equation}
	\dot{\rhoop} = \frac{1}{\im\hbar} [ \Hop, \rhoop] +\kappa \pares{\aop \rhoop \aopd-\frac{1}{2} \cpares{\aopd\aop,\rhoop}} \\-\frac{\Gamma}{2} \spares{\bopd + \bop,\spares{\bopd+\bop,\rhoop}},
\end{multline}
where $\rhoop$ is the density matrix of the cavity and the mechanical mode. The first dissipative term models cavity photon losses at a rate  $\kappa$ (i.e.  $\text{d}\mv{\aopd \aop}/\text{d} t = -\kappa \mv{\aopd \aop} + \dots$). The second dissipative term accounts for white mechanical displacement noise with a decoherence rate given by $\Gamma$ (i.e.  $\text{d}\mv{\bopd \bop}/\text{d} t = \Gamma + \dots$). This form of mechanical dissipation models laser recoil heating, scattering of air molecules in high vacuum, and any other type of displacement noise (e.g. trap vibrations) for levitated nanoparticles~\cite{GonzalezBallestero2019}. In~\cite{supplementary} we provide an analogous discussion for the standard mechanical dissipator describing the weak coupling to a thermal bath that is relevant for clamped mechanical oscillators. By writing the master equation \eqref{master-equation} in terms of the normal modes, and neglecting rapidly rotating terms (this can be done provided that $ \kappa \ll \Delta $), one can analytically obtain the asymptotic value of the minimal variance in the far-detuned regime $\Delta\gg \Omega$ (see~\cite{supplementary} for further details). The asymptotic variance is given by
\be \label{variance-dissipation}
\begin{split}
	\lim_{t r \gg 1} \varsq = \frac{\Omega}{2 \Delta } \pare{1 + \frac{\kappa}{4g}\sqrt{\frac{\Delta}{\Omega}} + \frac{\Gamma\Delta^2}{4g^3}\sqrt{\frac{\Delta}{\Omega}}   }.
\end{split}
\ee
The second and the third term represent the noise-induced correction to the minimal variance. Equation~\eqref{variance-dissipation} shows that in the presence of noise there is an optimal detuning $\Dopt$ for which the variance is minimized. The optimal detuning is well approximated by $\Dopt \approx g \sqrt{\kappa/(3 \Gamma)}$. The condition for unstable dynamics [\eqnref{unstable}] then reads $g/\Omega > \sqrt{\kappa/(48\Gamma)}$, and the condition for far detuning  [$\Dopt \gg \Omega$] reads $g/\Omega \gg \sqrt{3\Gamma/\kappa}$. Note that typically $\kappa \gg \Gamma$, especially when considering microcavities (see below for further details). Figure \ref{fig:dissipation} displays the asymptotic mechanical squeezing $S$  at optimal detuning as a function of the rates $\kappa/\Omega$ and $\Gamma/\Omega$  for different coupling rates $g/\Omega$. Figure \ref{fig:dissipation} shows that the generation of strong squeezing is feasible in the presence of cavity losses and mechanical displacement noise. Specifically, squeezing above 3 dB is possible even at $g/\Omega=1$, while strong squeezing ($S > 30 $ dB) can be achieved deep in the ultrastrong coupling regime $g/\Omega\gtrsim 100$. Note that  the resolved sideband regime, namely $\kappa \ll \Omega$, is not a requirement to obtain strong mechanical squeezing. 

\begin{figure}[t]
	\centering
	\includegraphics[width=1\linewidth]{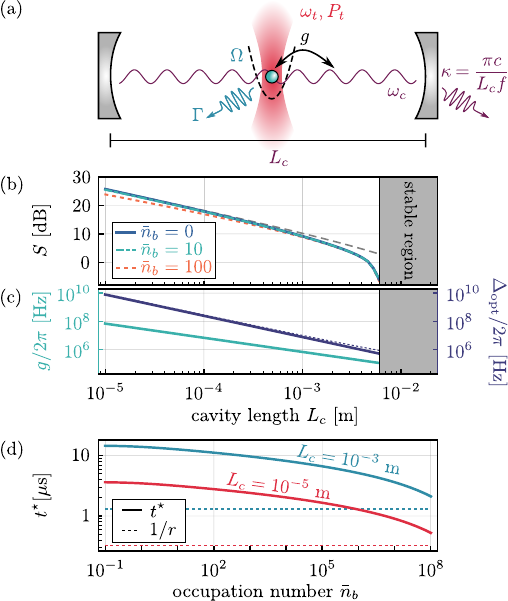}
	\caption{Feasibility of mechanical squeezing in coherent scattering. 
		(a) Schematic representation of the setup. A spherical nanoparticle is levitated by optical tweezers with power $P_t$, frequency $\omega_t$, and waist $W_t$. It is placed at a node of an optical cavity with length $L_c$, resonance frequency $\omega_c$, and decay rate $\kappa$. Optical tweezers provide a trapping potential with frequency $\Omega$, an effective  optomechanical coupling at a rate $g$, and particle recoil heating at a rate $\Gamma$.  
		(b) Asymptotic mechanical squeezing $S$ as a function of cavity length for a silica nanoparticle with radius $R = 100$ nm. The remaining parameters are $P_t = 29$ mW, $W_t = 0.7$ $\mu$m, $\lambda_t \equiv 2\pi c /\omega_t = 1064$ nm, $\lambda_c \equiv 2\pi c /\omega_c = 1064$ nm, $f=10^5$, and the asymmetry parameters of the tweezers $A_x=0.9$, $A_y = 0.8$ (see~\cite{supplementary}). 
		(c) Optomechanical coupling rate $g$ and optimal detuning $\Dopt$ as a function of cavity length, for the same parameters as in (b). Approximate and exact expressions for $\Dopt$ are shown by solid and dotted lines, respectively. 
		(d) Extension time $\tstar$ as a function of the initial mean phonon number, for the same parameters as in (b). 
	}
	\label{fig:feasibility}
\end{figure}

Let us now show that the results discussed above and displayed in \figref{fig:dissipation} are particularly feasible in levitated optomechanics via coherent scattering. We consider the setup shown schematically in \figref{fig:feasibility}(a), in which a nanoparticle is trapped in optical tweezers and placed at a node of an optical cavity. The scattering of laser photons into free space and into the cavity induces the coherent and dissipative dynamics that are well described by the master equation \eqref{master-equation} with the Hamiltonian given by \eqnref{Hamiltonian}~\cite{GonzalezBallestero2019}. Detailed expressions for the trapping frequency $\Omega$, the optomechanical coupling rate $g$, the recoil heating rate $\Gamma$, and the cavity photon decay rate $\kappa$ are given in~\cite{supplementary,GonzalezBallestero2019}. Here we discuss only their dependence on the most relevant quantities, that is, tweezers power $P_t$, particle radius $R$, cavity length $L_c$, and cavity finesse $f$. In particular,  $\Omega \propto P_t^{1/2}$, $g \propto P_t^{1/4}  R^{3/2} L_c^{-1}$, $\Gamma \propto P_t^{1/2} R^3 $, and $\kappa \propto L_c^{-1} f^{-1} $. Therefore, one obtains that $g^2/(\Omega  \Dopt) \propto R^3 L_c^{-1/2} f^{1/2} $. Hence, the instability condition \eqnref{unstable} at optimal detuning is independent of the tweezers power $P_t$ and benefits from high-finesse cavities with small mode volumes and large (but sub-wavelength) nanoparticles. 

The mechanical squeezing achievable in  coherent scattering is shown in \figref{fig:feasibility}(b). We plot the asymptotic mechanical squeezing $S$ as a function of the cavity length $L_c$ for initial mean phonon numbers $n_b =0, 10$, and 100, assuming a silica nanoparticle of radius $R=100$ nm, trap frequency $\Omega/(2\pi) = 100$ kHz, and a cavity finesse $f=10^5$ (see caption of \figref{fig:feasibility} for more details). As predicted, the smaller the cavity length, the larger the generated squeezing, with $S$ reaching values well over 10 dB for sub-millimeter-sized cavities~\cite{Hunger2010,Wachter2019,Fait2021}. The results shown in \figref{fig:feasibility}(b) are obtained by numerically solving the master equation \eqref{master-equation}, and they are compared with the analytical expression for the asymptotic variance \eqnref{AsymptoticX}, shown by the dashed gray line. Equation \eqref{AsymptoticX} is an excellent approximation away from the stable region (denoted by the shaded area) and for initial mean phonon numbers $\bar n_b \lesssim 100$. Figure \ref{fig:feasibility}(c) displays the optomechanical coupling rate $g$ and the optimal detuning $\Dopt$ corresponding to the case analyzed in~\figref{fig:feasibility}(b). The solid line shows the optimal detuning given by the approximation $\Dopt \approx g \sqrt{\kappa/(3\Gamma)}$, and it is nearly indistinguishable from the exact numerical value, shown by the dotted line. Note that the requirements $g,\Dopt \gg \Omega = 2 \pi \times 100~\text{kHz}$ are well satisfied. 

Since squeezing occurs at a given angle $\anglesq$~[see \figref{fig:unstable_dynamics}(a)], the spatial extension of the position probability distribution of the nanoparticle grows as a function of time. The linear dynamics of the nanoparticle in a coherent scattering setting is well described by the master equation \eqref{master-equation} and the Hamiltonian given in \eqnref{Hamiltonian} provided that this spatial extension is smaller than the cavity optical wavelength $\lambda_c$~\cite{GonzalezBallestero2019}. It is thus important to check that significant squeezing occurs before the spatial extension is too large. This is analyzed in \figref{fig:feasibility}(d), where we show the extension timescale $\tstar$, defined as the time such that $\spares{\mv{\Xop^2(\theta=0)  \hbar/{(m \Omega)}} (\tstar) }^{1/2} = 0.1 \lambda_c$, as a function of the initial mean phonon number $\bar n_b$ for two values of the cavity length (solid lines), and compare it to the squeezing timescale $1/r$ (dashed lines).  \figref{fig:feasibility}(d) confirms that the asymptotic value of $S$ can be achieved faster than $\tstar$ for a mechanical mode initially in a state given by a wide range of thermal occupation numbers even well above the ground state, evidencing the broad feasibility of squeezing. 

We remark that the setup considered here models recent coherent scattering experiments~\cite{Windey2019,Delic2019,Delic2020,Sommer2021,Ranfagni2021} that have demonstrated ground state cooling~\cite{Delic2020} and strong coupling~\cite{Sommer2021,Ranfagni2021}. We argue that the scheme we have presented is attainable in similar experiments where the cavity length is decreased, increasing the optomechanical coupling rate $g$ at the cost of increasing the cavity decay rate $\kappa$. Such small cavities with high finesse are experimentally feasible and have been realized in~\cite{Hunger2010,Wachter2019,Fait2021}. In this context, motional cooling can be achieved via feedback~\cite{Magrini2021,Tebbenjohanns2021} and the coupling to the microcavity can be switched on and off by controlling the detuning. In order to experimentally demonstrate the generation of mechanical squeezing we propose to optically measure the position of the particle during the stable harmonic dynamics generated by setting $\Delta \gg \Dopt$ at a time $t_0$ ($t^\star \gg t_0 \gg 1/r$), which effectively decouples the particle from the cavity. In particular, the recorded trajectories on different experimental runs can be demodulated at frequency $\Omega$ and angle $\phi\equiv -\anglesq-\pi/2$~\cite{Doherty2012,Rossi2019} and then ensemble averaged. As we show in~\cite{supplementary}, the variance of such a demodulated position operator is given by the squeezed variance \eqnref{variance-dissipation} at $t=t_0$ plus a noise term of the order of $\Gamma/\Omega$ that is much smaller than one. This method should thus allow one to experimentally measure the squeezed variance \eqnref{variance-dissipation}.

In summary, we have shown how a dynamical multimode instability in a linear optomechanical system can be used to rapidly generate strong mechanical squeezing. The instability can be exploited by operating the optomechanical system in the far red-detuned and ultrastrong coupling regime. Our results show that it is worth exploring the different types of dynamical instabilities encountered in a multimode system~\cite{Kustura2019} (e.g. three-mode dynamical instabilities might be used to generate two-mode entanglement via their separate coupling to a third mediating mode). While our results are in principle applicable to any optomechanical system, we have focused on an optically levitated nanoparticle coupled to a high-finesse optical cavity via coherent scattering. The combination of the recent achievement of free-space quantum control of nanoparticles~\cite{Magrini2021,Tebbenjohanns2021} with a properly designed microcavity opens a direct pathway towards  strong mechanical quantum squeezing induced by multimode instabilities. In this sense, our article provides a new use of optical cavities in the field of levitodynamics~\cite{GonzalezBallestero2021}, beyond passive cooling or mediating coupling between different particles~\cite{Rudolph2020,Brandao2021}: a microcavity is a great mechanical quantum squeezer.

We acknowledge valuable discussions with Uro\v{s} Deli\'{c}. This research was supported by the European Union’s Horizon 2020 research and innovation programme under grant agreement No. [863132] (IQLev) and by the Q-Xtreme project of the European Research Council under the European Union’s Horizon 2020 research and innovation program (grant agreement 951234).

%

\pagebreak
\onecolumngrid
\newpage
\begin{center} \textbf{\large Supplemental material} \end{center}
\newcommand{\beginsupplement}{
	\setcounter{table}{0}
	\setcounter{figure}{0}
	\setcounter{equation}{0}
	\setcounter{page}{1}
	\renewcommand{\thetable}{S\arabic{table}}
	\renewcommand{\thefigure}{S\arabic{figure}}
	\renewcommand{\theequation}{S\arabic{equation}}
	\renewcommand\thesection{S\arabic{section}}
}
\beginsupplement

The structure of the supplemental material is the following. In \secref{sec:T} we give the expression of the canonical transformation between the physical modes $\aop,\bop$ and the normal modes $\cop_1,\cop_2$. In \secref{sec:master-equation} we summarize the derivation of an approximate master equation for the normal mode $\cop_2$ and the asymptotic value of the minimal variance of the mechanical mode. In \secref{sec:optomechanical} we discuss the generation of mechanical squeezing in the presence of a standard mechanical dissipator describing weak coupling to a thermal bath. In \secref{sec:coherent-scattering} we provide detailed expressions for all coherent and dissipative rates in a coherent scattering setup. In \secref{sec:detection} we describe in detail the proposed squeezing detection scheme.

\section{Canonical transformation into a normal form}\label{sec:T}
The Hamiltonian Eq. (1) in the main text can be transformed into its normal form by a canonical transformation $T$, which relates the physical operators $\Psopvec \equiv (\aop,\bop,\aopd,\bopd)^T$ and the normal mode operators $\Phopvec \equiv (\cop_1,\cop_2,\copd_1,\copd_2)^T$ as $\Phopvec = T^{-1} \Psopvec$. It reads~\cite{Kustura2019}
\be\label{T}
T = G^{\dagger} P G,
\ee
with
\be \label{T-parts}
\begin{split}
	G &= \frac{1}{\sqrt{2}} 
	\begin{pmatrix} 
		\mathbb{1}_2 & \mathbb{1}_2 \\  
		-\im\mathbb{1}_2 & \im\mathbb{1}_2   
	\end{pmatrix},\\
	P &= \begin{pmatrix}
		\frac{ \bfac{+}}{ \sqrt{\afac{+}} }\pare{\frac{\Delta}{\zeta} } & 
		-\bfac{-}^2  \pare{\frac{\Delta^2}{2g \Omega} } &  
		0  &  
		\frac{1}{\afac{-}} \pare{\frac{g \Omega}{\zeta^2} } \\
		\frac{2}{\bfac{+} \sqrt{\afac{+}} } \pare{\frac{g \Omega}{\Delta \zeta} }&
		1 & 
		0 & 
		\frac{-2}{\afac{-}\bfac{-}^2} \pare{\frac{g \Omega}{\Delta \zeta}}^2\\
		0 & 
		\frac{1}{\afac{-}}\spare{\bfac{-}^2 \pare{\frac{\Delta^2}{2 g \Omega}}
			-\frac{2g}{\Delta} }& 
		\frac{1}{\afac{+}^{3/2}} \spare{ 
			\bfac{+}\pare{\frac{\Delta}{\zeta}}
			+ \frac{4}{\bfac{+} } \pare{\frac{g^2\Omega}{\Delta^2 \zeta}} }& 
		\frac{1}{\afac{-}^2} \spare{ \frac{g \Omega}{\zeta^2}
			- \frac{4}{\bfac{-}^2} \pare{\frac{g^3 \Omega^2}{\Delta^3 \zeta^2}} }\\
		0 & 
		\afac{-}\pare{\frac{\Delta}{ \Omega } }& 
		\frac{2\sqrt{\afac{+}}}{\bfac{+}} \pare{\frac{g}{\zeta}} & 
		\frac{2}{\bfac{-}^2} \pare{\frac{g^2 \Omega}{\Delta \zeta^2}}\\
	\end{pmatrix},
\end{split}
\ee
where $\mathbb{1}_2$ is the two-dimensional identity matrix, $\zeta^2 \equiv \sqrt{(\Delta^2-\Omega^2)^2+ 16 \Delta \Omega g^2}$, and we define the dimensionless factors
\be \label{definitions}
\begin{split}
	\afac{\pm} \equiv \sqrt{ \frac{\zeta^2 \pm \pare{\Delta^2 +\Omega^2}}{2 \Delta^2} }, \quad
	\bfac{\pm} \equiv \sqrt{ \frac{\zeta^2 \pm \pare{\Delta^2 -\Omega^2}}{2 \Delta^2} }.
\end{split}
\ee
The canonical transformation \eqnref{T} fulfills the property $T^\dagger \pseudo T = T \pseudo T^\dagger = \pseudo$, where $\pseudo \equiv \text{diag} (\mathbb{1}_2,-\mathbb{1}_2)$, and as a consequence it preserves the bosonic commutation relations, namely $[\psop_i,\psop_j^\dagger] = [\phop_i,\phop_j^\dagger] = \pseudo_{ij} $. In the far-detuned regime $\Delta \gg \Omega$, \eqnref{T-parts} can be approximated as 

\be \label{T-simplified}
\begin{split}
	P & \overset{}{\approx} \begin{pmatrix}
		1 & -\frac{2 g}{\Delta} &  0  &  \frac{1}{2} \sqrt{\frac{\Omega}{\Delta}}\\
		0 & 1 & 0 & - \frac{\Delta}{4 g} \sqrt{\frac{\Omega}{\Delta}}\\
		0 & -\frac{4g^2}{\Delta^2}\sqrt{\frac{\Omega}{\Delta}}  & 1 & 0 \\
		0 & \frac{2 g}{\sqrt{\Delta\Omega} } & \frac{2g}{\Delta} & \frac{1}{2} \\
	\end{pmatrix}.
\end{split}
\ee

\section{Approximate master equation for the squeezed normal mode}\label{sec:master-equation}
We apply \eqnref{T} to the master equation [Eq. (6) in the main text] in order to derive an approximate master equation where the dynamics of the two normal modes are uncoupled. The effect of \eqnref{T} on the coherent part of the dynamics is that it transforms the Hamiltonian into its normal form [Eq. (3) in the main text]. Let us analyze the transformation of the incoherent part, given by the dissipators 
\be \label{dissipator}
\begin{split}
	\mathcal{D}_\text{cavity}[\rhoop] + \mathcal{D}_\text{particle}[\rhoop] \equiv \kappa \pares{\aop \rhoop \aopd-\frac{1}{2} \cpares{\aopd\aop,\rhoop}}-\frac{\Gamma}{2} \spares{\bop+\bopd,\spares{\bop+\bopd,\rhoop}} . 
\end{split}
\ee
Using \eqnref{T} it can be shown that the cavity dissipator in terms of the normal modes $\Phopvec \equiv (\cop_1,\cop_2,\copd_1,\copd_2)^T$ reads $\mathcal{D}_\text{cavity}[\rhoop] = \sum_{i,j=1}^4 T_{1i} T_{1,j}^*\pares{\phop_i \rhoop \phop_j^\dagger-\frac{1}{2} \cpares{\phop_j^\dagger\phop_i,\rhoop}}$, and it assumes the most general form for quadratic Lindblad dissipators, which includes terms describing absorption, decay, incoherent two-mode squeezing, and incoherent two-mode interaction between the normal modes. However, 
the coupling between different modes can be neglected in the regime $\kappa \ll \Delta$, where these terms are far off resonance. This leads to a dissipator in which the two normal modes $\cop_1,\cop_2$ are uncoupled. Specifically, the contribution to the squeezed $\cop_2$-mode dynamics reads 
\be\label{dissipator-c}
\begin{split}
	\mathcal{D}_\text{cavity}^{(c_2)}[\rhoop] =  \gamma^{\text{(d)}} \mathcal{L}_{\cop_2,\copd_2}[\rhoop] + \gamma^{\text{(a)}} \mathcal{L}_{\copd_2,\cop_2}[\rhoop]
	+ \pare{  w \mathcal{L}_{\cop_2,\cop_2}[\rhoop]  + \hc },
\end{split}
\ee 
with $\mathcal{L}_{\iop_1, \iop_2}[\rhoop]=\iop_1 \rhoop \iop_2-\cpares{\iop_2\iop_1,\rhoop}/2$ for arbitrary operators $\iop_{1,2}$, and the rates $\gamma^\text{(d)} = \kappa \vert  T_{12}\vert^2$, $\gamma^\text{(a)} = \kappa \vert  T_{14}\vert^2$, $w = \kappa T_{12} T_{14}^*$. The latter can be simplified in the regime $\Delta \gg \Omega$ using \eqnref{T-simplified}, leading to $\gamma^\text{(d)} = \gamma^\text{(a)} \approx \kappa g^2/\Delta^2 $ and $w \approx \kappa g^2/\Delta^2 +\im \kappa r/{(4\Delta)}$, with the squeezing rate $r\approx 2 g \sqrt{\Omega/\Delta}$. 
The particle dissipator can be analyzed in an analogous way. Using \eqnref{T} we have $\mathcal{D}_\text{particle}[\rhoop] = -\Gamma \sum_{i,j=1}^{4} P_{2i} P_{2j} \spares{\Sop_i,\spares{\Sop_j,\rhoop}}$, where $\Sopvec \equiv G \Phopvec$, with $P$ and $G$ defined in \eqnref{T-parts}. In the regime $\kappa \ll \Delta$ we can neglect the off-resonant terms, leading to a dissipator in which the two normal modes $\cop_1,\cop_2$ are uncoupled. Specifically, the contribution to the $\cop_2$-mode dynamics reads 
\be\label{dissipator-p}
\mathcal{D}_\text{particle}^{(c_2)}[\rhoop] = -\frac{\Gamma}{2} \cpare{ \vphantom{\frac{1}{2}}
	\spares{\cop_2+\copd_2,\spares{\cop_2+\copd_2,\rhoop}}
	- P_{24}^2 \spares{\copd_2-\cop_2,\spares{\copd_2-\cop_2,\rhoop}}
	+  \pare{ 4 \im P_{24}\vphantom{\aopd}  \mathcal{L}_{\cop_2,\cop_2} [\rhoop] + \hc } 
}.
\ee

Together with the normal form Hamiltonian [Eq. (3) in the main text], Eqs. (\ref{dissipator-c}-\ref{dissipator-p}) lead to a closed system of equations
\be \label{eom-n}
\begin{split}
	\fd{}{ t} \begin{pmatrix} \mv{\copd_2\cop_2} \\ \mv{\copd_2 {}^2}  \\\mv{\cop_2^2}	\end{pmatrix} = 
	\begin{pmatrix}
		0 & r & r \\	
		2r & 0 & 0 \\	
		2r & 0 & 0 \\	
	\end{pmatrix} 
	\begin{pmatrix} \mv{\copd_2\cop_2} \\ \mv{\copd_2 {}^2}  \\\mv{\cop_2^2}	\end{pmatrix}
	+
	\begin{pmatrix} \gamma^\text{(a)} +\Gamma \vert\eta\vert^2  \\ r-w-\Gamma \eta^2\\ r-w^*-\Gamma(\eta^*)^2	\end{pmatrix},
\end{split}
\ee
with $\eta = 1 - \im P_{24} \approx 1 + \im {\sqrt{\Delta\Omega}}/{(4g)} $. The initial conditions for the dynamics are obtained from the physical initial conditions using \eqnref{T}. Consequently, \eqnref{eom-n} can be solved, leading to analytical expressions for the second moments of the squeezed mode, $\mv{\copd_2\cop_2}(t)$, $\mv{\copd_2 {}^2}(t)$, $\mv{\cop_2^2}(t)$. 

The minimal mechanical variance $\varsq$ can be obtained by expressing the mechanical second moments $\mv{\bopd\bop}(t)$ and $\mv{\bop^2}(t)$ in terms of the normal mode second moments using \eqnref{T}. Here, all the moments $\mv{\phop_i \phop_j}$ with $i,j=1,3$ can be neglected, as they depend on the unsqueezed $\cop_1$-mode, and therefore have a negligible effect on $\varsq$. This allows us to express $\varsq$ as an analytical function of time in terms of $\mv{\copd_2\cop_2}(t)$, $\mv{\copd_2 {}^2}(t)$, and  $\mv{\cop_2^2}(t)$. Finally, the asymptotic value $\lim_{t r \gg 1} \varsq$ [given by Eq. (7) in the main text] can be obtained as the zeroth order term in the series expansion of $\varsq$ in powers of $x\equiv e^{-r t/\Delta}$.

\section{Squeezing in the presence of a thermal optomechanical dissipator}\label{sec:optomechanical}
\begin{figure}[t]
	\centering
	\includegraphics[width=0.5\linewidth]{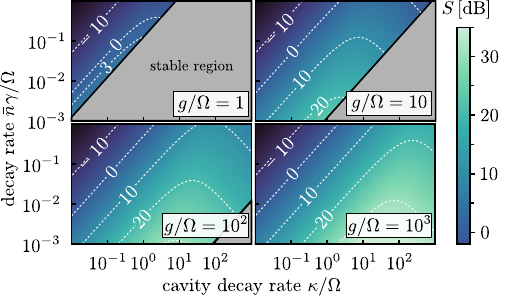}
	\caption{Asymptotic mechanical squeezing $S$ in the presence of thermal optomechanical dissipator, considering the detuning $\Dopt \approx \Omega \kappa / (2\bar n \gamma)$, and an initial state with mean phonon number $\bar n_b = 0$.  }
	\label{fig:optomechanical}
\end{figure}

Let us consider the dynamics in the presence of a standard thermal optomechanical dissipator, which is relevant for clamped mechanical oscillators. The optomechanical dynamics in this case is described by the master equation
\be \label{ME-OM}
\begin{split}
	\fd{\rhoop}{t} = 
	&-\frac{\im}{\hbar} [\Hop,\rhoop] + \kappa \mathcal{L}_{\aop, \aopd}[\rhoop]+\gamma (1+\bar n) \mathcal{L}_{\bop, \bopd}[\rhoop] + \gamma \bar n \mathcal{L}_{\bopd, \bop}[\rhoop].
\end{split}
\ee 
Here $\rhoop$ is the density operator of the total system, $\mathcal{L}_{\iop, \iopd}[\rhoop]=\iop \rhoop \iopd-\cpares{\iopd\iop,\rhoop}/2$ for an arbitrary operator $\iop$, $\kappa$ is the cavity decay rate, $\gamma$ is the decay rate of the mechanical mode due to the weak coupling to a thermal bath, and $\bar n = \spares{\exp\pares{\hbar \Omega / k_B T}-1}^{-1}$ is the occupation of the bath at the oscillator frequency. Using \eqnref{T}, \eqnref{ME-OM} can be written in terms of the normal modes $\Phopvec \equiv (\cop_1,\cop_2,\copd_1,\copd_2)^T$. We can neglect the coupling between different modes in the regime  $\kappa, \bar n \gamma \ll \Delta$, where these terms are off-resonant. This leads to a dissipator in which the two normal modes $\cop_1,\cop_2$ are uncoupled, and the dynamics of the squeezed $\cop_2$-mode is described by an approximate master equation which can be solved analytically, in an analogous way as discussed in \secref{sec:master-equation}. Making an inverse transformation and neglecting contributions from the unsqueezed $\cop_1$-mode, we obtain the asymptotic value of the minimal variance of the mechanical mode. In the far red-detuned regime, namely $\Delta\gg \Omega$, it reads
\be \label{variance-OM}
\begin{split}
	\lim_{t r \gg 1} \varsq = \frac{\Omega}{2 \Delta } \spare{1 + \frac{\kappa}{4g}\sqrt{\frac{\Delta}{\Omega}} + \frac{\bar n \gamma}{2 g} \pare{\frac{\Delta}{\Omega}}^{3/2}   }. 
\end{split}
\ee
\eqnref{variance-OM} is minimized by the detuning $\Dopt \approx \Omega \kappa / (2\bar n \gamma)$, and the resulting asymptotic squeezing is shown in \figref{fig:optomechanical} as a function of dissipation rates and for different coupling rates $g/\Omega$. 

\section{Details on levitated cavity optomechanics via coherent scattering}\label{sec:coherent-scattering}

Here we provide detailed expressions for the trapping frequency $\Omega$, the optomechanical coupling rate $g$, the recoil heating $\Gamma$, and the cavity photon decay rate $\kappa$ in a specific setup for optical levitodynamics. Namely, we consider an optically levitated particle coupled to an optical cavity through coherent scattering. The rates appearing in the particle-cavity interaction Hamiltonian [Eq. (1) in the main text] read $\Delta = \omega_c-\omega_t$, where $\omega_c$ and $\omega_t$ are the cavity mode frequency and the tweezers frequency, respectively, and~\cite{GonzalezBallestero2019}
\be \label{rates-H}
\begin{split}
	\Omega &= \frac{2}{A_y^2 W_t^2} \pare{\frac{ P_t \alpha}{\pi \varepsilon_0 c \hspace{0.5mm}  m}  \frac{A_y}{A_x}  }^{\frac{1}{2}},\quad
	g      = \frac{1}{c W_c} \spare{\frac{P_t}{4 c \hspace{0.5mm}m  L_c^2 } \frac{A_y}{A_x} \pare{\frac{\alpha \omega_c^2}{\pi \varepsilon_0}}^3}^\frac{1}{4}.
\end{split}
\ee 
Here $W_t$ is the tweezers waist, $A_{x,y}$ are dimensionless factors which account for the asymmetry of the transverse beam profile (with $A_x = A_y = 1 $ denoting the symmetric case), and $P_t$ is tweezers power; $m$ is the nanoparticle mass and $\alpha = 3\varepsilon_0 V (\varepsilon - 1)/(\varepsilon + 2)$ is the polarizability, with the nanoparticle volume $V$, the vacuum permittivity $\varepsilon_0$ and the relative permittivity $\varepsilon$; $L_c$ is the cavity length, $W_c = \sqrt{c L_c / \omega_c}$ is the cavity waist for a confocal cavity, and $c$ is the vacuum speed of light. The decoherence rates appearing in the particle-cavity effective master equation [Eq. (6) in the main text] are given by~\cite{GonzalezBallestero2019}
\be \label{recoil}
\begin{split}
	\kappa &= \frac{\pi c }{ L_c f},\quad
	\Gamma = \frac{\omega_t^2}{30 c^5} \spare{ \frac{P_t}{ c \hspace{0.5mm}m}\frac{A_y}{A_x}\pare{\frac{\alpha \omega_t^2}{\pi \varepsilon_0}}^3}^{\frac{1}{2}},
\end{split}
\ee 
with the cavity finesse $f$.

\section{Detection of squeezing}\label{sec:detection} 

\begin{figure}[t]
	\centering
	\includegraphics[width=1\linewidth]{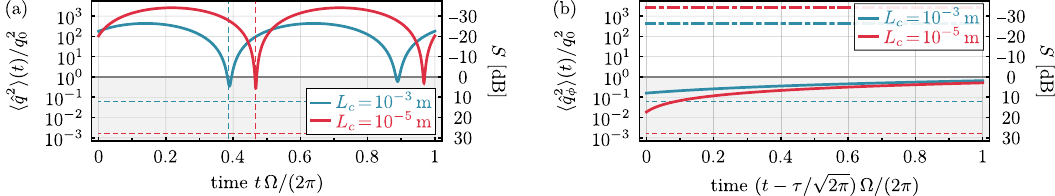}
	\caption{(a) Position variance $\mv{ \hat{q}^2 }(t)$ and (b) demodulated variance $\mv{ \hat{q}_\phi^2 }(t)$ for $\phi = -\anglesq -\pi/2$ (solid lines) and $\phi = -\anglesq$ (dot-dashed lines) as a function of time during step (ii) of the protocol. For the initial condition we consider a squeezed mechanical state generated during step (i) from the state $\bar n_b=0$ by the unstable dynamics inside a microcavity of length $L_c$. We set $ t_0 = 3/r$, $\tau=2\pi /\Omega$, and the values of the remaining parameters are the same as in the caption of Fig.~3 in the main text. Vertical dashed lines in (a) indicate the time at which the minimal position variance is achieved. Horizontal dashed lines indicate the asymptotic minimal variance obtained during step (i) [Eq.~(7) in the main text]. Gray shaded area indicates squeezing.}
	\label{fig:detection}
\end{figure}

Here we analyze a measurement scheme to detect mechanical squeezing generated in the unstable regime [Eq.~(2) in the main text]. Our protocol consists of two steps. In step (i), we let the system evolve in the unstable regime during time $t \in [-t_\text{0},0)$ (with $t^\star \gg t_0 \gg 1/r$). During this step, squeezing at a phase-space angle $\anglesq$ is generated as discussed in the main text, and the variance $\lim_{t_0 r \gg 1} \Delta^2 X_\text{sq}$ [Eq. (7) in the main text] is achieved. In step (ii), at the time $t=0$ we set the cavity detuning to $\Delta\gg\Delta_\text{opt}$, which effectively decouples the cavity and the mechanical mode. The dynamics of the mechanical mode at $t>0$ are thus given by a free evolution in the harmonic trap.  As we show below, mechanical squeezing prepared during step (i) can be extracted by demodulation of a position measurement during step (ii)~\cite{Doherty2012,Rossi2019}. 

The dynamics during step (ii) is described by the following master equation, 
\be\label{master-equation-ii}
\dot{\rhoop}_b = -\im\Omega [ \bopd\bop, \rhoop_b] -\frac{\Gamma}{2} \spares{\bopd + \bop,\spares{\bopd+\bop,\rhoop_b}}, 
\ee
where $\rhoop_b$ is the density matrix of the mechanical mode. The initial state for the dynamics in \eqnref{master-equation-ii} is the squeezed state obtained in step (i) using a cavity of length $L_c$. Equation~\eqref{master-equation-ii} leads to a closed system of equations for the second moments which can be  solved. From the continuous position measurement of the particle in this step, one can extract the time evolution of the variance of the position operator, defined as $\hat{q} \equiv \sqrt{2}\hspace{0.5mm}q_0\Xop(\theta=0)=q_0(\bop+\bopd)$, with $q_0 \equiv \sqrt{\hbar/(2 m \Omega)}$. This position variance as a function of time is shown in \figref{fig:detection}(a). As expected, one observes that the variance changes as a function of time due to the rotation of the squeezed state in phase space induced by the evolution in the harmonic potential. At the time $\Omega t_\text{min} \equiv \theta_\text{sq} + \pi/2$ [denoted in \figref{fig:detection}(a) by vertical dashed lines], the position variance reaches a minimal value that is given by 
\be
\frac{\mv{\hat{q}^2}(t_\text{min})}{2 q_0^2} =\lim_{t_0 r \gg 1} \Delta^2 X_\text{sq} + \frac{\Gamma}{2\Omega}\spares{2 \theta_\text{sq} + \pi + \sin (2\theta_\text{sq})}.
\ee
While squeezing below zero-point motion (indicated by the gray shaded area) can still be observed, the value is larger than $\lim_{t_0 r \gg 1} \Delta^2 X_\text{sq}$ due to the displacement noise experienced during the time $t_\text{min}$ required to rotate in phase space until the minimal variance is in the position operator. 

In order to observe larger squeezing, comparable to the one generated after step (i), one could perform a demodulation of each position trajectory before the average is performed. That is, one is interested in a demodulated position operator defined by 
$\hat q_\phi (t) \equiv 2 \int_{-\infty}^{\infty}  \hat q(t')\cos(\Omega t' + \phi) f_\tau(t'-t) dt'$. 
Here the function $f_\tau(t)$ describes the demodulation window and is given by $f_\tau(t) = \exp (- t^2/\tau^2)/(\tau \sqrt{\pi})$. The demodulation time window $\tau$ is assumed to be larger than the oscillation period of the particle, namely $\tau \gg 2 \pi /\Omega$. From the average of the recorded demodulated trajectories, one has access to the variance $\mv{ \hat{q}_\phi^2 }(t)$ during step (ii), that is 
\be \label{int2}
\begin{split}
	\mv{ \hat{q}_\phi^2 }(t) =  4\integral{-\infty}{\infty}{s'}\integral{-\infty}{\infty}{s''} \mv{\hat{q}(s')\hat{q}(s'')} 
	\cos(\Omega s' + \phi) \cos(\Omega s'' + \phi) f_\tau(s'-t)f_\tau(s''-t).
\end{split}
\ee 
The correlation functions $\mv{\hat{q}(s')\hat{q}(s'')} $ can be obtained using the quantum regression theorem~\cite{Carmichael}, leading to
\be \label{q-var-int}
\begin{split}
	\mv{ \hat{q}_\phi^2 }(t) &=  8q_0^2\integral{-\infty}{\infty}{s'}\integral{-\infty}{\infty}{s''}g(s',s'')   
	\cos(\Omega s' + \phi) \cos(\Omega s'' + \phi) f_\tau(s'-t)f_\tau(s''-t),
\end{split}
\ee
with 
\be \label{g}
\begin{split}
	g(s',s'')   \equiv  
	\frac{1}{2} e^{-\im \Omega(s'-s'')} + \mv{\bopd\bop}(0) &\cos\spare{\Omega(s'-s'')} + \text{Re} \cpare{\spare{\mv{\bop^2}(0) -\frac{\im \Gamma}{2\Omega}} e^{-\im \Omega(s'+s'')}} \\
	&+2\Gamma \Theta(s'-s'') \spare{ s'' \cos\spare{\Omega(s'-s'')} + \frac{1}{2\Omega} \sin\spare{\Omega(s'-s'')}},
\end{split}
\ee
where $\Theta(x)$ is the Heaviside step function. Equation~\eqref{q-var-int} can be evaluated analytically using the Sokhotski-Plemelj identity to define $\Theta(x)$ in terms of its Fourier transform. The final result can be significantly simplified in the regime $\tau \gg 2 \pi /\Omega$, and we obtain for the demodulated variance 
\be \label{q-var}
\begin{split}
	\frac{\mv{ \hat{q}_\phi^2 }(t)}{2 q_0^2} &= \mv{\Xop^2(- \phi)}(0) + \frac{\Gamma}{2\Omega} \sin(2\phi) + \Gamma \pare{t - {\frac{\tau}{\sqrt{2\pi  } }}  }.  
\end{split}
\ee
Equation~\eqref{q-var} is well defined in the regime $t>\tau/\sqrt{2\pi}$, which ensures that the demodulation window, described by the function $f_\tau(t)$, lies within the regime corresponding to step (ii), namely $t>0$. 

Setting the demodulation phase to $\phi = -\anglesq-\pi/2$ allows us to extract the asymptotic minimal variance reached in the unstable regime, since $\mv{\Xop^2(\anglesq+\pi/2)}(0) \equiv \lim_{r t_0\gg 1}\varsq$. Note that the retrieval of the original minimal variance prepared in step (i) is in general prevented due to the displacement noise. However, as demonstrated by time evolution of $\mv{ \hat{q}_{-\anglesq-\pi/2}^2}$ in \figref{fig:detection}(b) (solid lines), significant levels of squeezing can nevertheless be achieved, as compared with the unmodulated case in \figref{fig:detection}(a). For comparison, in \figref{fig:detection}(b) we show the time evolution of $\mv{ \hat{q}_{-\anglesq}^2}$ (dot-dashed lines), corresponding to the anti-squeezed quadrature. This further emphasizes the squeezing achieved for $\phi = -\anglesq-\pi/2$ (indicated by the gray shaded area). Demodulation of a position measurement during free evolution in the harmonic trap thus offers a viable route for the detection and characterization of mechanical squeezing generated in the unstable regime.  

\end{document}